\documentstyle[12pt]{article}

\newcommand{\be}{\begin{equation}}
\newcommand{\ee}{\end{equation}}
\newcommand{\ba}{\begin{eqnarray}}
\newcommand{\ea}{\end{eqnarray}}
\begin{document}
%%%%%%%%%%%%%%%%%%%%%%%%%%%%%%%%%%%%
%Without pictures use this macro
%\def\pct#1{(see Fig. #1.)}
%%%%%%%%%%%%%%%%%%%%%%%%%%%%%%%%%%%
%With pictures use this macro
%\def\pct#1{\input epsf \centerline{ \epsfbox{#1.eps}}}

%%%%%%%%%%%%%%%%%%%%%%%% FRONT PAGE %%%%%%%%%%%%%%%%%%%%%%%%%%%%%%%%%%%%%
\begin{titlepage}
\hbox{\hskip 12cm ROM2F-97/40  \hfil}
\hbox{\hskip 12cm November 1997}
\vskip .8cm
\begin{center}  {\Large  \bf Aspects of Type I Compactifications\\
and Type I - Heterotic Duality}
 
\vspace{1.2cm}
 
{\large Carlo Angelantonj}

\vspace{.3cm}

{\it Dipartimento di Fisica Universit{\`a} de L'Aquila \\
INFN -- LNGS \\ 
Via Vetoio \ 67010 \ Coppito (AQ) \ \ ITALY\\
and\\
Dipartimento di Fisica, Universit\`a di Roma ``Tor Vergata''\\
Via della Ricerca Scientifica, 1, 00133  Roma ITALY\\
angelantonj@aquila.infn.it}

\end{center}

\vskip .6cm

\abstract{We review the construction of open descendants of the
type IIB superstring on the $Z$-orbifold. It results in a chiral 
four-dimensional model with gauge group $SO(8) \otimes U(12)$ and three 
generations of matter in the $(8,12^*)\oplus (1,66)$ representations.
As a test of type I - heterotic duality, that reduces to a weak/weak duality
in $D=4$, a heterotic model on the same orbifold is also presented. The 
massless spectrum reproduces exactly the one found in the type I case 
apart from
additional twisted matter charged with respect to the $SO(8)$ gauge group.
The puzzle is solved by noting that at generic points in the moduli space
these states get masses.}

\vfill \end{titlepage}
%%%%%%%%%%%%%%%%%%%%%%%%%%%%%%%%%%%%%%%%%%%%%%%%%%%%
\makeatletter
\@addtoreset{equation}{section}
\makeatother
\renewcommand{\theequation}{\thesection.\arabic{equation}}
\addtolength{\baselineskip}{0.3\baselineskip} 
%%%%%%%%%%%%%%%%%%%%%%%%%%%%%%%%%%%%%%%%%%%%%%%%%%%%

\section{Introduction}

During the last few years, 
conjectures have been proposed
establishing relations between apparently different
string theories in various dimensions
\cite{wittd,duag,dualities}. 

In string perturbation theory
there are two topologically distinct classes of 
theories: those with only oriented closed strings (type IIA and B, 
heterotic $E_8\otimes E_8$ and $SO(32)$) and those with both open and closed 
unoriented strings (type I).
The picture that is recently taking shape suggests  that these five theories
describe different regions of an underlying moduli space associated
to a more fundamental theory yet to be discovered. 
The transition functions between the 
different regions are realized in terms of duality transformations. In 
many cases these dualities require the inversion of the coupling constant, i.e.
they are inherently non-perturbative. This explains why different theories
look completely different in perturbation theory.

In order to check and then extract informations
from dualities a special role
is  played by BPS states and, in particular, by type II solitons
charged with respect to  the 
Ramond-Ramond (RR) fields \cite{polch}, called D-branes. 
A D-brane represents the manifold where open string ends are free to move, 
thus giving a 
geometrical interpretation of Chan-Paton (CP) multiplicities.
A microscopic description in terms of open strings with Dirichlet boundary 
conditions in the directions transverse to the D-brane world-volume has 
brought about
remarkable progress not only in the string duality context
but also for what concerns (nearly extremal) black-holes thermodynamics.
This is the reason why recently there has been an increasing interest in type I 
strings, although the initial proposal of identifying open 
string theories as {\it parameter space orbifolds} of left-right symmetric 
theories of oriented closed strings \cite{cargese} 
was already fully and consistently systematized some time ago \cite{bs,comp}. 
Type I vacua have been analyzed in some detail in several dimensions
\cite{bs,sixd,gepm,bl,ks,chiral,cp}, resulting in new interesting phenomena.
For instance one finds a
rich pattern of CP symmetry breaking and varying numbers
of tensor multiplets, including zero \cite{gepm}, in $D=6$. 
Other puzzling phenomena 
occur at the boundary of the moduli space, as for example the appearance of 
tensionless strings \cite{aug,tens}.

\section{Type I - Heterotic Duality}

One example of string-string duality is given by the two ten-dimensional 
theories with gauge group $SO(32)$: the type I and heterotic strings 
\cite{wittd}. At the massless level these theories certainly agree. 
Moreover, the low energy 
effective action
is uniquely fixed by $N=(1,0)$ supersymmetry.
Might they in 
fact be equivalent? Moreover, the agreement is such that strong 
coupling of one theory would turn into weak coupling of the other.
This is an essential point in any possible comparison between the
two theories, since weak 
coupling of one is certainly not equivalent to weak coupling of the other.
Denoting by $g_{I(H)}$ and $\phi_{I(H)}$ the ten-dimensional metric and dilaton 
of type I (heterotic) string, the duality map reads:
\be
g_{I} = e^{-\phi_H} g_H\,, \qquad \phi_{I}=-\phi_{H} \,.
\ee
In support of this string-string duality, it can be shown that 
the excitations of 
the type I D-string exactly coincide with the light-cone degrees of freedom of 
the $SO(32)$ heterotic string \cite{pw}. The inverse relation seems harder to 
establish since the type I string is unoriented and carries no conserved 
charge.

Upon toroidal compactification the map between the two theories gets much more 
involved. In particular the relation between the heterotic and type I dilatons 
in $D$ dimensions is \cite{chiral}
\be
\phi^{(D)}_{I} ={6-D\over 4} \phi_{H}^{(D)} - {(D-2) \over 16} \log \det
G_{H}^{(10-D)}\,, \label{dilatons}
\ee
where $G^{(10-D)}_{H}$ is the internal metric 
in the heterotic string frame, and 
there is a crucial sign change at $D=6$. From this relation
one can deduce that there 
always exists a region in the moduli space where both type I and heterotic 
string theories are weakly coupled, and there we can rely on perturbation 
theory, which we understand. 

\section{$D=6$ Type I Vacua}

The first consistent $D=6$ $N=1$ chiral open-string models \cite{bs} 
differ markedly from perturbative 
heterotic K3 compactifications \cite{walton}, since they include 
different numbers of tensor multiplets 
that take part in a generalized Green-Schwarz mechanism \cite{aug}. 
In the last two years, additional 
instances of $N=1$ type I $D=6$ models have been constructed 
as irrational toroidal orbifolds \cite{sixd}.
For rational internal tori there is an elegant description of 
superstring propagation on manifolds of 
$SU(n)$ holonomy in terms of tensor products of $N=2$ 
superconformal minimal models \cite{gep}.
In addition to the Virasoro generators, $L_n$, the $N=2$ 
superconformal algebra includes two 
supercurrents $G^{\pm}_{r}$ and a $U(1)$ current, $J_n$. 
An important feature of the $N=2$
superconformal algebra is the presence of an automorphism, 
known as spectral flow, that connects
different sectors of the spectrum. The minimal models form a discrete 
series with central charge $c_k
= 3k/(k+2)$. Gepner has shown how to construct $D$-dimensional 
string vacua with space-time
supersymmetry tensoring $N=2$ minimal models in such a way 
that the total internal central charge
is $c_I = \sum_i c_{k_i} = 12 - 3 (D-2)/2$, 
where $c_{k_i}$ are the central charges of the various
factors \cite{gep,eoty}.

In six dimensions there are several possible types of Gepner models. 
For the type IIB K3 compactifications
the chiral spectrum is uniquely fixed by target-space 
$N=(2,0)$ supersymmetry and anomaly cancellation, 
and results in a supergravity multiplet coupled to 21 
tensor multiplets \cite{romans}. The scalar fields of the 
resulting low-energy supergravity parameterize the coset 
$SO(5,21)/SO(5)\otimes SO(21)$.

The 
general construction of perturbative open string vacuum configurations consists 
in a non-geometrical ${\bf Z}_2$-orbifold, named parameter space orbifold or 
orientifold \cite{cargese,ps} (see also \cite{ikprev} for more details). 
First of all, the conventional Polyakov perturbative series must 
be supplemented with the inclusion of world-sheets with boundaries and/or 
crosscaps. The truncation of the parent left-right symmetric closed string 
spectrum encoded in the torus partition function, ${\cal T}$, is implemented by 
the Klein bottle projection, ${\cal K}$. As a result the $Z_2$ projection
halves the number of space-time supersymmetries.
These two contributions make up the 
``untwisted sector'' of the parameter space orbifold. The role of the 
``twisted sector'' is played by the open string spectrum encoded in the 
annulus partition function, ${\cal A}$, and its projection, the M\"obius
strip, ${\cal M}$. In standard geometrical orbifolds twisted sectors have 
multiplicities associated to the fixed points. Similarly, in parameter space 
orbifolds, the open string states may acquire multiplicities associated to 
their ends through the introduction of CP factors, or multiple D-branes.
Consistency requirements may be deduced transforming the above amplitudes to 
the transverse channel, where Klein bottle, $\tilde {\cal K}$, annulus,
$\tilde{\cal A}$, and M\"obius strip, $\tilde{\cal M}$, amplitudes are related
to tree level closed string amplitudes between boundary and/or crosscap states, 
and consist in the cancellation of tadpoles of unphysical massless states or,
equivalently, the cancellation of total RR charge.

Then in $D=6$ the open descendants have $N=(1,0)$ 
target-space supersymmetry, and the closed
unoriented spectrum consists of the supergravity 
multiplet coupled to $n^{c}_{T}$ tensor
multiplets and $n^{c}_{H}$ hypermultiplets. 
The uniqueness of the parent type IIB massless
spectrum forces $n^{c}_{T} + n^{c}_{H} =21$, 
since the Klein bottle projection simply halves the 
fermionic degrees of freedom. The open unoriented 
spectrum completes the construction of the
parameter space orbifold consistently with 
anomaly cancellation. Actually, tadpole conditions are
in one-to-one correspondence with anomaly cancellation. 
For details on the various models
and the corresponding CP gauge groups we refer the reader 
to the original paper \cite{gepm}.
Let us just quote the model $D_{81}$ in \cite{gepm}. 
It is an open descendant of the type IIB superstring
compactified on six copies of Gepner minimal models with $k=1$. 
The peculiarity of this model
is that the closed unoriented spectrum {\it does not} contain 
any tensor multiplet at all. It is very
interesting because the heterotic dual model should 
correspond to a vacuum configuration with a frozen dilaton!

It is worth to stress that typically the rank of the
CP gauge group is smaller than the one related
to irrational orientifolds, and this is due 
to the presence of a non-vanishing background for the
NS-NS antisymmetric tensor in the internal tori 
\cite{bps}.

The relation (\ref{dilatons}) implies that dual (non-perturbative) 
heterotic vacua in $D=6$ should
correspond to orbifold compactifications in which the 
usual modular invariance constraints are violated \cite{afiuv}.

\section{A Non-Trivial Dual Pair in $D=4$}

Now let us discuss the first class of four dimensional type I chiral models, 
discovered in \cite{chiral}. We start from the $Z$-orbifold of the 
type IIB superstring. Then, the twist is given by $({1\over 3} ,{1\over 3} ,
{1\over 3} )$, where each factor acts on a $T^2$ torus, whose metrics
$G_{ab} = {1\over 2} R^2 C_{ab}$, is 
proportional to the 
$SU(3)$ Cartan matrix $C_{ab}$ \cite{orbs}, . 
Moreover, we choose a vanishing NS-NS antisymmetric 
tensor in order to get a CP gauge group of maximal size \cite{bps}.

The action of the orbifold point group breaks the ten-dimensional $SO(8)$ 
characters down to $SO(2) \otimes SU(3) \otimes U(1)$ ones. The torus amplitude 
can then be written as
\ba   
{\cal T} &=& {1 \over 3} \Xi_{0,0}(q) \Xi_{0,0}(\bar q) \sum q^{p_{L}^{2}/2}
\bar q ^{p_{R}^{2}/2} + {1 \over 3} \sum_{\epsilon = \pm 1} 
\Xi_{0,\epsilon}(q)  \Xi_{0,\epsilon}(\bar q)
\nonumber \\
& & + {1 \over 3} \sum_{\eta = \pm 1} \sum_{\epsilon =0,\pm 1} 
\Xi_{\eta,\epsilon}(q) \Xi_{-\eta,-\epsilon}(\bar q) \,,
\label{torus}
\ea  
where we have introduced
\ba
\Xi_{0,\epsilon}(q) &=& \left( { A_0 \chi_0 + \omega^\epsilon A_+ \chi_- + 
\bar\omega ^\epsilon A_- \chi_+  \over H_{0,\epsilon}^{3}} \right) (q) \,,
\nonumber \\
\Xi_{\pm,\epsilon}(q) &=& \left( { A_0 \chi_\pm + 
\omega^\epsilon A_\pm \chi_0 +  \bar\omega ^\epsilon A_\mp \chi_\mp  \over
H_{\pm,\epsilon}^{3}} \right) (q) \,,
\label{chars}
\ea 
and $\{ A_0 , A_+, A_- \}$ are supersymmetric characters of conformal weights 
$\{ {1\over 2}, {1\over 6}, {1\over 6}\}$ respectively, $\{ \chi_0 , \chi_+ ,
\chi_- \}$ are level-one $SU(3)$ characters of conformal weights 
$\{ 0, {1\over 3}, {1\over 3}\}$ respectively, and
\ba
H_{0,\epsilon}(q)  &=&  q^{1/12}
\prod_{n=1}^{\infty} (1-\omega^\epsilon q^n)  (1 - \bar\omega ^\epsilon q^n) \,,
\nonumber \\
H_{+,\epsilon}(q) &=& H_{-,-\epsilon}(q) = 
{1\over\sqrt{3}} q^{-1/36}
\prod_{n=0}^{\infty} (1 - \omega^\epsilon q^{n+1/3}) (1 -  \bar\omega 
^\epsilon q^{n+2/3}) \,,
\label{lattice}
\ea
originate from the action of the twist on the bosonic coordinates.
Moreover the left and right momenta are given by
$(p_a )_{L,R} = m_a \pm {1\over 2} G_{ab} n^b$.

The massless spectrum then comprises the $N=2$ supergravity multiplet coupled 
to $9+1$ hypermultiplets from the untwisted sector and 27 
hypermultiplets from the twisted sectors, one for each fixed point. 

The torus partition function (\ref{torus}) corresponds to the charge 
conjugation modular invariant. Together with the fact that the only real 
character in the $Z$-orbifold is the identity, the Klein bottle amplitude gets 
contributions only from the untwisted sector, and its expression is given by
\be  
{\cal K} = {1 \over 6} \Xi_{0,0}(q^2) \sum_{m_a}   q^{m_a G^{ab} m_b}
+ {1 \over 6} \Xi_{0,+}(q^2) + {1 \over 6} \Xi_{0,-}(q^2)  \,.
\label{klein}
\ee  
It contains only the conventional sum over the momentum 
lattice since, for generic values
of the internal tori volumes, the condition 
$p_L = \omega p_R$ (where $\omega$ is the $Z_3$ generator) 
does not have any non-trivial
solutions. 
In the open sector this reflects the presence
of only D9-branes.
The massless states in the projected closed string spectrum comprise the 
$N=1$ supergravity multiplet coupled to $1+9+27$ chiral multiplets. 
The scalars 
in the $1+9$ untwisted chiral multiplets parametrize the K\"ahler manifold
$Sp(8,{\bf R})/SU(4)\times U(1)$, a real slice of the coset manifold 
$E_{6(+2)} /SU(2) \times SU(6)$ parameterized by the untwisted scalars in the 
parent type IIB theory \cite{scalars}.

The twisted sector of the parameter space orbifold, to be identified with the 
open string spectrum, contains the annulus and M\"obius strip amplitudes:
\ba   
{\cal A} &=& {(N+M+\bar M)^2 \over 6 } 
\Xi_{0,0}(\sqrt{q}) 
\sum_{m_a}  q^{m_a G^{ab} m_b} 
\nonumber \\    
& & + {(N +\omega M + \bar\omega \bar M )^2 \over 6} 
\Xi_{0,+}(\sqrt{q}) +
{(N +\bar\omega M + \omega \bar M )^2 \over 6} 
\Xi_{0,-}(\sqrt{q}) \,,
\label{annulus}
\ea   
\ba  
{\cal M} &=& - {(N+M+\bar M ) \over 6}  \hat\Xi _{0,0} ( -\sqrt{q} )
\sum_{m_a} q^{m_a G^{ab} m_b}
\nonumber \\  
& & - {(N+\bar\omega M +\omega\bar M ) \over 6} \hat\Xi _{0,+} ( - \sqrt{q} ) 
-  {(N +\omega M + \bar\omega \bar M ) \over 6} \hat\Xi _{0,-} ( - \sqrt{q} )
\,,
\label{moebius}
\ea 
where $N,\, M,\, \bar M$ are CP multiplicities. 
The M\"obius amplitude presents some subtleties connected with the proper
definition of a set of real ``hatted'' characters \cite{bs}.
Tadpole cancellations in the 
transverse channel result in
\ba   
N + M + \bar M &=& 32
\nonumber \\  
N - {1 \over 2} (M + \bar M ) &=& -4  \,.
\label{tadpole}
\ea   
From the amplitudes (\ref{annulus}), (\ref{moebius}) and 
tadpole conditions (\ref{tadpole}), 
we can extract the CP gauge group and the massless charged 
matter. In particular we have
\be
G_{CP} = SO(8) \otimes SU (12) \otimes U(1)\,,
\ee
with three generations of chiral multiplets in the $({\bf 8}_v , {\bf 12}^* )_{
-1} \oplus ({\bf 1}, {\bf 66})_{+2}$ representations. The cancellation of the
twisted tadpole guarantees that this chiral spectrum is anomaly free, aside 
from the $U(1)$ factor. The $U(1)$ anomaly translates into a Higgs-like 
mechanism that gives the abelian vector a mass of the the string scale
\cite{dsw}.

The candidate heterotic dual  corresponds to a
perturbative compactification on the $Z$-orbifold with non-standard embedding
\cite{chiral}. 
The action of the twist on the gauge degrees of freedom
consists of four copies of the basic ${\bf Z}_3$ 
twist $({1\over 3}, {1\over 3} ,{1\over 3})$ and  
breaks the ten-dimensional $SO(32)$ 
gauge group down to $SO(8) \otimes U(12)$, the CP group of the type I model.
Moreover, the untwisted charged spectrum coincides with the open string 
spectrum of the type I model, {\it i.e.} chiral multiplets in the 
representations $({\bf 8}_v , {\bf 12}^* ) \oplus ({\bf 1},{\bf 66})$.
A striking feature of the heterotic model is that twisted scalars are charged 
with respect to the gauge group; in fact, in the heterotic case, we get 27 
additional chiral multiplets in the $({\bf 8}_c , {\bf 1})$ representation of 
the surviving (non-anomalous) gauge group, $SO(8) \otimes SU(12)$.
The apparent puzzle associated to the presence of these extra charged chiral 
multiplets can be solved if one analyses the perturbative superpotentials for 
the type I and heterotic string models just discussed \cite{kaku}.

Denoting by $\Phi^{is}_{a}$ and $\chi^{k}_{[rs]}$ the 
three generations of chiral multiplets in the 
$({\bf 8}_v , {\bf 12}^*)$ and $({\bf 1},{\bf 66})$ representations of the 
gauge group, the cubic superpotential of the type I model is 
fixed by gauge symmetry and global $SU(3)$ symmetry to be \cite{chiral}
\be  
W_I = y_I \ \delta^{ab} \ \epsilon_{ijk} \ \Phi^{ir}_{a} \Phi^{js}_{b} 
\chi^k_{[rs]} \,.
\ee  
In the heterotic model the perturbative superpotential is more involved 
since the scalar fields from the twisted sector have non-trivial 
couplings. Denoting by $S^A$ the 27 blow-up modes and by $T^{B}_{\dot\alpha}$ 
the 27 twisted charged scalars in 
the $({\bf 8}_c , {\bf 1})$ representation, the
heterotic superpotential reads \cite{kaku}
\be
W_H = y_H \ \delta^{ab} \ \epsilon_{ijk} \ \Phi^{ir}_{a} \Phi^{js}_{b} 
\chi^k_{[rs]} 
+ \Lambda_{ABC} \delta^{\dot\alpha \dot\beta} S^A T^{B}_{\dot\alpha}
T^{C}_{\dot\beta} \,.
\ee
Notice that the additional contribution to the superpotential
is non-vanishing only for $A=B=C$ or $A\not=B\not=C\not=A$, where
$A,B,C$ label the 27 fixed points. The contribution coming from
states sitting at different fixed points is exponentially
suppressed with respect to the separation of the fixed points.
The form of $W$ and the above considerations suggest 
the solution of the puzzle. After blowing-up the 
orbifold singularities on the heterotic model (which amounts  giving  a 
non-vanishing vev to the $S^A$ fields) the $T$ fields become massive and 
decouple from the spectrum \cite{kaku}.
As a result, the type I and heterotic vacua are perturbatively equivalent.

\section{Discussion}

The advent of string dualities has shed some 
light on non-perturbative aspects of string
theories and supersymmetric Yang-Mills theories. 
In particular, type I - heterotic duality
seems  very fruitful in understanding 
non-perturbative effects in $D=4$ $N=1$
supersymmetric theories. A strong/weak 
coupling duality in $D=10$, reduces
after compactification to a perturbative 
duality in $D=4$. Nevertheless,
studying heterotic duals of $D=4$ type I vacua with 
D5-branes \cite{ks,cp} could help us
to learn about  non-perturbative effects in the 
heterotic string theory (such as NS
5-brane dynamics and the generation of a 
non-perturbative superpotential) by mapping
them onto perturbative effects on the type I side 
(such as D5-brane dynamics).

A deeper understanding of the relation between 
tadpoles and anomaly cancellations in 
$D=4$ type I vacua might also shed some light 
on some puzzling phenomena that occur
in four-dimensional orientifolds \cite{cp}.

\vskip 24pt
\begin{flushleft} {\large \bf Acknowledgments}
\end{flushleft}

I would like to thank the organizers for the kind invitation and 
G. Preparata and S.-J. Rey for discussions.
It is a pleasure to thank M. Bianchi, G. Pradisi,
A. Sagnotti and Ya.S. Stanev for the stimulating collaboration. 
This work was supported in part by CNR-KOSEF bilateral agreement.

\end{document}